\documentclass{ws-p8-50x6-00p}

\def\NPB#1#2#3{Nucl.~Phys.~B {\bf #1} (19#2) #3}
\def\PLB#1#2#3{Phys. Lett. B {\bf #1} (19#2) #3}

\def\PRD#1#2#3{Phys. Rev. D {\bf #1} (19#2) #3}

\def\ov{\overline}

\begin{document}

\title{Chiral Type II Orientifold Constructions 
%constructions
\\ as
M Theory on $G_2$ holonomy spaces\footnote{SUSY'01 proceedings contribution}}

\author{Mirjam Cveti\v c and Gary Shiu}

\address{Department of Physics and Astronomy, University of Pennsylvania,\\
Philadelphia, PA 19104, USA \\
E-mail: cvetic@dept.physics.upenn.edu, shiu@dept.physics.upenn.edu}

\author{Angel M.~Uranga}

\address{Departamento de F\'{\i}sica Te\'orica C-XI
and Instituto de F\'{\i}sica Te\'orica  C-XVI,\\
Universidad Aut\'onoma de Madrid,
Cantoblanco, 28049 Madrid, Spain \\
E-mail: angel.uranga@cern.ch}

\pubnum{UPR-968-T\\IFT-UAM-CSIC-01-36\\FT-UAM-01-21}

\maketitle

\abstracts{We summarize some recent progress in constructing
four-dimensional supersymmetric chiral models from Type II 
orientifolds.
%\cite{CSU1,CSU2}.
 We present the construction a supersymmetric Standard-like
Model and a supersymmetric GUT model
to illustrate the new features of this approach and  its connection to 
M  theory  on compact, singular $G_2$ holonomy spaces. 
The Standard-like model presented is the first example 
of a three-family supersymmetic orientifold model
with the Standard Model
as part of the gauge structure. 
We also discuss the connection of how chiral fermions arise in
this class of models with recent
results of M theory compactified on $G_2$
holonomy spaces.\\}

\section{Introduction}

If string theory is relevant to Nature, at low energies it should give
rise to an effective theory containing the Standard Model. 
Whether string theory can live up to
this promise depends on {\it how} the string vacuum describing the observable
world is selected among a highly degenerate moduli space -- 
a question that
we know very little about. Nevertherless, one can use experimental
constraints as guiding principles to construct semi-realistic
models, and explore with judicious assumptions, 
the resulting physical implications, in
particular to particle physics.
This is the basic premise of {\it string phenomenology} -- hopefully, by
exploring the generic features of string derived models, we can learn
some new stringy
physics that are important for low
energy predictions.

Until a few years ago, the
construction of four-dimensional string theory solutions 
 was carried out mainly
in the framework of weakly coupled heterotic string, 
in which a number
of semi-realistic models have been constructed and analyzed\cite{review}.
Meanwhile, model building from other string theories did not seem very
promising, partly due to the no-go theorem of 
perturbative Type II strings\cite{TypeIInogo}.
However, M theory unification made important 
 progress in uncovering non-perturbative
aspects of string theory: we now understand that these perturbative  models represent
only a corner of M theory -- the true string vacuum may well be in
a completely different regime in which the perturbative
heterotic string description breaks down.

Our view of string phenomenology changed drastically with the advent of
D-branes. The techniques of conformal field theory in describing
D-branes and orientifold planes allow, in principle for the 
construction of semi-realistic
string models in another calculable regime of M theory, 
as illustrated by the various
four-dimensional $N=1$ supersymmetric Type II 
orientifolds\cite{ABPSS,berkooz,N1orientifolds,zwart,ShiuTye,afiv,wlm,CPW,kr,CUW}.
In these models, chiral fermions appear on the worldvolume of the
D-branes since they are located at orbifold singularities in the internal
space. Semi-realistic models from non-supersymmetric type IIB orientifolds
of this kind have been constructed in \cite{aiq,aiqu}, with supersymmetry
breaking due to the presence of brane-antibrane configurations in the
model \cite{adsau}.

An alternative to obtain chiral fermions, which has only recently been
exploited in model building is to consider branes at angles.
In certain configurations of intersecting D-branes, the spectrum of open
strings stretched between them may contain chiral fermions, localized at
the intersection\cite{bdl}. This fact was employed 
in \cite{bgkl,afiru,bkl,imr,bonn,bklo} in the construction of
non-supersymmetric brane world models. However, the dynamics to determine
the stability of non-supersymmetric models are not well understood.
The purpose of \cite{CSU1,CSU2} is to explore orientifold models
with branes at angles that preserve ${\cal N}=1$ supersymmetry in
four dimensions.

This general class of supersymmetric orientifold models corresponds in
the strong coupling limit to M theory compactification on purely
geometrical backgrounds admitting a $G_2$ holonomy metric. We also discuss the
relation of this work with the recent results about the appearance of
four-dimensional chiral fermions in compactifications
of M theory on singular $G_2$ holonomy spaces \cite{AW,Witten,aW,CSU1,CSU2}.

\section{Model Building from Orientifolds and Intersecting D6-branes}

In this section we shall provide the key features of the construction.
We refer the readers to the original paper\cite{CSU1,CSU2} for 
more detailed discussions.

For concreteness, we consider an orientifold of type IIA on ${\bf
T}^6/({\bf Z}_2\times {\bf Z}_2)$. Generalization to other orbifolds would
involve similar techniques, and presumably analogous final results. The
orbifold actions have generators $\theta$, $\omega$ acting as $ \theta:
(z_1,z_2,z_3) \to (-z_1,-z_2,z_3)$, and  $\omega: (z_1,z_2,z_3) \to
(z_1,-z_2,-z_3)$ on the complex coordinates $z_i$ of ${\bf T}^6$, which is 
assumed to be factorizable. The orientifold action  is $\Omega R$, where
$\Omega$ is world-sheet parity, and $R$ acts by $ R:\  (z_1,z_2,z_3) \to
({\overline z}_1,{\overline z}_2,{\overline z}_3)$. The model contains
four kinds of O6-planes, associated to the actions of $\Omega R$, $\Omega
R\theta$, $\Omega R \omega$, $\Omega R\theta\omega$. The cancellation of
the RR crosscap tadpoles requires an introduction of $K$ stacks of $N_a$
D6-branes ($a=1,\ldots, K$) wrapped on three-cycles (taken to be the
product of 1-cycles $(n_a^i,m_a^i)$ in the $i^{th}$ two-torus), and their
images under $\Omega R$, wrapped on cycles $(n_a^i,-m_a^i)$. In the case
where D6-branes are chosen parallel to the O6-planes  (orientifold 6-planes), the resulting model
is related by T-duality to the orientifold in \cite{berkooz}, and is
non-chiral. Chirality is however achieved using D6-branes intersecting at non-trivial
angles.

The cancellation of untwisted tadpoles impose constraints on the number of
D6-branes and the types of 3-cycles that they wrap around. The
cancellation of twisted tadpoles determines the the orbifold actions on
the Chan-Paton indices of the branes (which are explicitly given in
\cite{CSU1,CSU2}).
%For each stack of D6$_a$-branes, and
%their $\Omega R$ images, denoted by D6$_{a'}$-branes, the actions are as
%follows:
%\begin{eqnarray}
%&& \gamma_{\theta,a} =  \diag(i \id_{N_a/2},-i \id_{N_a/2}\, ; 
%-i \id_{N_a/2},i \id_{N_a/2}) \nonumber \\
%\nonumber \\
%&& \gamma_{\omega,a}  =  \diag \left[ 
%\pmatrix{0 & \id_{N_a/2} \cr -\id_{N_a/2} & 0 } \; ; \;
%\pmatrix{0 & \id_{N_a/2} \cr -\id_{N_a/2} & 0 } \right] \nonumber \\
%\nonumber \\
%&& \gamma_{\Omega R,a}  = 
%\pmatrix{ & & \id_{N_a/2} & 0 \cr 
%& & 0 & \id_{N_a/2} \cr 
%\id_{N_a/2} & 0 & & \cr 
%0 & \id_{N_a/2} & & \cr }
%\end{eqnarray}
%(zzz Is it useful to give these matrices?)

The condition that the system of branes preserves ${\cal N}=1$
supersymmetry
requires \cite{bdl} that each stack of D6-branes  is related to
the O6-planes by a rotation in $SU(3)$: denoting by $\theta_i$ the angles
the D6-brane forms with the horizontal direction in the $i^{th}$
two-torus, supersymmetry preserving configurations must
satisfy
$
\theta_1\, +\, \theta_2\, +\, \theta_3\, =\, 0
$.
This in turn impose a constraint on the wrapping numbers and the
complex structure moduli $\chi_i=R_2^i/R_1^i$.

The rules to compute the spectrum are analogous to those in \cite{bkl}.
Here, we summarize the resulting chiral spectrum in Table \ref{matter}
found in \cite{CSU1,CSU2},
where
\begin{equation}
I_{ab}\ =\
(n_a^1m_b^1-m_a^1n_b^1)(n_a^2m_b^2-m_a^2n_b^2)(n_a^3m_b^3-m_a^3n_b^3)
\label{internumber}
\end{equation}
\medskip
\begin{table}[htb] \footnotesize
\renewcommand{\arraystretch}{1.25}
\begin{center}
\begin{tabular}{|c|c|}
\hline {\bf Sector}   & 
{\bf Representation} \\
\hline\hline
$aa$    &  \hspace{1cm} $U(N_a/2)$ vector multiplet \\
       & \hspace{1cm} 3 Adj. chiral multiplets  \\
\hline\hline
$ab+ba$   &  $I_{ab}$ chiral multiplets in
$(N_a/2,\overline{N_b/2})$ rep.   \\
\hline\hline
$ab'+b'a$ &  $I_{ab'}$ chiral multiplets in $(N_a/2,N_b/2)$ rep.
  \\
\hline\hline
$aa'+a'a$ &  $-\frac 12 (I_{aa'} - \frac{4}{2^{k}} I_{a,O6})$
chiral multiplets in sym. rep. of $U(N_a/2)$  \\
          & $-\frac 12 (I_{aa'} + \frac{4}{2^{k}} I_{a,O6})$ 
chiral multiplets in antisym. rep. of $U(N_a/2)$ \\
\hline
\end{tabular}
\end{center}
\caption{\small General spectrum on D6-branes at generic angles
(namely, not parallel to any O6-plane in all three tori). The spectrum is
valid for tilted tori. The models may contain additional non-chiral pieces
in the $aa'$ sector and in $ab$, $ab'$ sectors with zero intersection, if
the relevant branes overlap.
\label{matter} }
\end{table}

\subsection{\bf Standard-Like Model}

Let us present an example leading to a three-family
Standard-like Model
massless spectrum. The D6-brane configuration is provided in table \ref{cycles3family},
and satisfies the tadpole cancellation conditions. 
The configuration is supersymmetric for
$\chi_1:\chi_2:\chi_3=1:3:2$.

\begin{table} 
%\footnotesize
[htb] \footnotesize
\renewcommand{\arraystretch}{1.25}
\begin{center}
\begin{tabular}{|c||c|l|}
\hline
Type & $N_a$ & $(n_a^1,m_a^1) \times
(n_a^2,m_a^2) \times (n_a^3,\widetilde{m}_a^3)$ \\
\hline
$A_1$ & 8 & $(0,1)\times(0,-1)\times (2,{\widetilde 0})$ \\
$A_2$ & 2 & $(1,0) \times(1,0) \times (2,{\widetilde 0})$ \\
\hline
$B_1$ & 4 & $(1,0) \times (1,-1) \times (1,{\widetilde {3/2}})$ \\
$B_2$ & 2 & $(1,0) \times (0,1) \times (0,{\widetilde {-1}})$ \\
\hline
$C_1$ & 6+2 & $(1,-1) \times (1,0) \times (1,{\widetilde{1/2}})$ \\
$C_2$ & 4 & $(0,1) \times (1,0) \times (0,{\widetilde{-1}})$ \\
\hline
\end{tabular}
\end{center}
\caption{\small D6-brane configuration for the three-family model.}
\label{cycles3family}
\end{table}

The resulting spectrum
is given in table \ref{spectrum3}, where the last column provides the charges
under a particular anomaly-free $U(1)$ linear combination which plays the
role of hypercharge. The spectrum of chiral multiplets, regarding their
quantum numbers under the Standard Model group $SU(3)\times SU(2)\times
U(1)_Y$, corresponds to three quark-lepton generations, plus a number of
vector-like Higgs doubles, as well as an anomaly-free set of chiral exotic
matter. This last set of states is chiral under the Standard Model group,
so it cannot be made massive until electroweak symmetry breaking. Hence
the model suffers from the presence of light exotics which most likely
render it unrealistic. Our main point in presenting it is, however, to
illustrate the possibility of building semirealistic models in our setup.
It is conceivable that one can construct more realistic 
models with this approach
such that these phenomenological problems are absent.

\begin{table} 
%\footnotesize   
[htb] \footnotesize
\renewcommand{\arraystretch}{1.25}
\begin{center}
\begin{tabular}{|c|r|c|c|c|c|c|c|}
\hline 
Sector & 
%$U(3)\times U(2)\times USp(2)^2\times USp(4)$ &
Non-Abelian Reps. &
$Q_3$ & $Q_1$ & $Q_2$ & $Q_8$ & $Q_8'$ & $Q_Y$ 
%& Field 
\\
\hline
$A_1 B_1$  & 
$3 \times 2\times (1,{\ov 2},1,1,1)$ & 
0 & 0 & $-1$ & $\pm 1$ & 0 & $\pm \frac 12$ 
%& $H_U$, $H_D$
\\
          & 
$3\times 2\times (1,{\ov 2},1,1,1)$ &
0 & 0 & $-1$ & 0 & $\pm 1$ & $\pm \frac 12$ 
%& $H_U$, $H_D$
\\
$A_1 C_1$ & 
$2 \times (\ov{3},1,1,1,1)$ &
$-1$ & 0 & 0 & $\pm 1$ & 0 & $\frac 13, -\frac 23$ 
% & $U$, $D$
\\
         & 
$2 \times (\ov{3},1,1,1,1)$ &
$-1$ & 0 & 0 & 0 & $\pm 1$ & $\frac 13, -\frac 23$ 
%& $U$, $D$
\\
          & 
$2 \times (1,1,1,1,1)$ &
0 & $-1$ & 0 & $\pm 1$ & 0 & $1,0$ 
%& $E$, $\nu_R$
\\
          & 
$2 \times (1,1,1,1,1)$ &
0 & $-1$ & 0 & 0 & $\pm 1$ & $1,0$ 
%& $E$, $\nu_R$
\\
$B_1 C_1$ & 
$(3,{\ov 2},1,1,1)$ &
1 & 0 & $-1$ & 0 & 0 & $\frac 16$
% & $Q_L$
\\
            & 
$(1,{\ov 2},1,1,1)$ &
0 & 1 & $-1$ & 0 & 0 & $-\frac 12$ 
%& $L$
\\
$B_1 C_2$ & 
$(1,2,1,1,4)$ &
0 & 0 & $1$ & 0 & 0 & 0 
%& 
\\
 $B_2 C_1$ & 
$(3,1,2,1,1)$ &
1 & 0 & 0 & 0 & 0 & $\frac 16$ 
%& 
\\
          & 
$(1,1,2,1,1)$ &
0 & 1 & 0 & 0 & 0 & $-\frac 12$ 
%&
\\
$B_1 C_1^{\prime}$ & 
$2\times (3,2,1,1,1)$ &
1 & 0 & 1 & 0 & 0 & $\frac 16$ 
%& $Q_L$
\\ 
                   & 
$2\times (1,2,1,1,1)$ &
0 & 1 & 1 & 0 & 0 & $-\frac 12$ 
%& $L$ 
\\
$B_1 B_1^{\prime}$ & 
$2\times (1,1,1,1,1)$ & 
0 & 0 & $-2$ & 0 & 0 & 0 
%& 
\\
                   & 
$2\times (1,3,1,1,1)$ & 
0 & 0 & $2$ & 0 & 0 & 0 
%& 
\\ \hline
\end{tabular}
\end{center}
\caption{\small Chiral Spectrum of the open string sector in the
three-family model. The non-Abelian gauge group is
$SU(3) \times SU(2) \times USp(2) \times USp(2) \times USp(4)$.
Some vector-like sectors have not been included for the sake of clarity.
%zzz I have removed this: Notice that we have not included the $aa$ sector
%piece, even though it is generically present in the model. The non-chiral
%pieces in the $ab$, $ab'$ and $aa'$ sectors are not present for branes at
%generic locations, hence they are not listed here. 
\label{spectrum3}}
\end{table}

\subsection{\bf GUT Model}

Here we present a string model leading to a four-family $SU(5)$ grand
unified theory. The D6-brane configuration is

\begin{center}
\begin{tabular}{|c|c|}
\hline
$N_a$  & $(n_a^1,m_a^1)\times (n_a^2,m_a^2)\times (n_a^3,m_a^3)$
\\
\hline
$10+6$  & $(1,1)\times (1,-1) \times (1,{\widetilde {1/2}})$ \\
\hline
$16$ & $(0,1)\times (1,0) \times (0,-{\widetilde {1}})$ \\
\hline
\end{tabular}
\end{center}
which is supersymmetric for $\arctan \chi_1 - \arctan \chi_2 + \arctan
(\chi_3/2)=0$. We consider that the first set of $16$ branes is split in
two parallel stacks of $10$ and $6$. The resulting spectrum is
\begin{eqnarray}
& U(5)\times U(3) \times USp(16) & \nonumber \\
& 3(24+1,1,1) + 3(1,8+1,1) + 3(1,1,119+1) & \nonumber \\
& 4({\ov{10},1,1}) + (5,1,16) + 4 ({\ov 5},{\ov 3},1) + (1,3,16) +
4(1,3,1)& \end{eqnarray}
The model is a four-family $SU(5)$ GUT, with additional gauge groups and
matter content. Notice that turning on suitable vev's for the adjoint
multiplets the model corresponds to splitting the $U(5)$ branes.
This provides a geometric interpretation of the GUT Higgsing to
the Standard Model group upon splitting $U(5)\to U(3)\times
U(2)\times U(1)$.

\section{Relation to  compastification of M theory on $G_2$ holonomy spaces}

M theory compactification on a manifold $X$ with $G_2$ holonomy gives rise
to an ${\cal N}=1$ theory in four dimensions. If $X$ is smooth, the low
energy theory is relatively uninteresting since it contains (in addition
to ${\cal N}=1$ supergravity) only abelian vector multiplets and neutral
chiral multiplets.\footnote{On the other hand, 
smooth special holonomy spaces provide
suitable backgrounds  for constructing regular fractional brane configurations
as viable gravity duals of strongly coupled field theories. For a review, see
\cite {cglp} and references therein.}  However, non-Abelian gauge symmetries 
and chiral
fermions can arise when the manifold $X$ is singular. Isolated conical
singularities of $G_2$ manifolds have been studied recently from different
points of view \cite{acharya,AMV,AW,Witten,aW,CSU1,CSU2}. We now discuss 
how some of these results can be understood  within the  orientifold setup.

A useful  approach to building  $G_2$ holonomy spaces is to construct type IIA
configurations preserving four supercharges and lifting them to M theory.
However, not all four-dimensional ${\cal N}=1$ supersymmetric vacua from M
theory correspond to $G_2$ holonomy compactifications, since the M theory lifts may 
may contain additional sources, i.e.  M-branes or G-fluxes,  other than a pure
gravitational background. Hence one needs to start with IIA configurations
containing D6-branes, O6-planes (and/or RR 1-form backgrounds, which are absent in 
our setup), only.  When lifted to M theory, a collection of $N$ D6-branes becomes
a multi-centered Taub-NUT space \cite{townsend}, whereas an O6-plane
becomes an Atiyah-Hitchin manifold \cite{SeibergWitten}. Hence, 
IIA configurations involving these ingredients, and preserving
four supercharges, when lifted to M theory correspond to a purely geometrical
background, {\it i.e.}, of 11 dimensional space-time compactified on 
a $G_2$ holonomy space. In this respect, the models considered here
correspond  to  M theory compactified on $G_2$ holonomy space which give rise to
non-Abelian
gauge symmetries and chiral fermions. The origin of the non-Abelian gauge
symmetries is well-known: gauge bosons arises from the massless M2-brane
states wrapped in the collapsed 2-cycles in the multi-Taub-NUT lift of
overlapping IIA D6-branes \cite{Sen}. In the following we remark on the
appearance of chiral fermions  from the M theory viewpoint.

In configurations where the RR 7-form charges are locally cancelled
(namely, 2 D6-branes and their 2 images are located on top of each
O6-plane), the M theory lift is remarkably simple. The M theory circle is
constant over the base space $B_6$, leading to a total variety $(B_6\times
S^1)/{\bf Z}_2$, where the ${\bf Z}_2$ flips the coordinate parametrizing
the M theory circle, and acts on $B_6$ as an antiholomorphic involution
(hence changing the holomorphic 3-form to its conjugate). This is the type
of configurations considered in\cite{blumen4d,bonn1,kmg} and the resulting
models are non-chiral.

In models with D6-branes at angles, chiral fermions arise. In fact, the
type IIA description with intersecting D6-branes allows to identify 
the nature of the singularities of the  $G_2$ holonomy space which 
lead to chiral fermions.
The following
analysis also makes contact with \cite{aW}. Away from the intersections of IIA
D6-branes and/or O6-planes, the IIA configuration corresponds to D6-branes
and O6-planes wrapped on (disjoint) smooth supersymmetric 3-cycles, which
we denote generically by $Q$. The corresponding $G_2$  holonomy space hence
corresponds to fibering a suitable hyperkahler four-manifold over each
component of $Q$. That is, an $A$-type ALE singularity for $N$ overlapping
D6-branes, and a $D$-type ALE space for D6-branes on top of
O6-planes (with the Atiyah-Hitchin manifold for no D6-brane, and its
double covering for two D6-branes etc., as follows from
\cite{SeibergWitten}).
Intersections of objects in type IIA therefore lift to co-dimension
7-singularities, which are isolated up to orbifold singularities.
%Since the gauge bosons are
%localized on a collection of D6-branes wrapping around a supersymmetric
%three-cycle, the corresponding $G_2$ manifold can be written locally as
%some ADE singularities fibered over a 3-manifold $Q$, where $Q$ is
%essentially a supersymmetric 3-cycle except at points where the branes
%intersect. 
It is evident from the IIA picture that the chiral fermions are localized at
these singularities. 

The structure of these singularities has been studies in \cite{aW}, and is
similar to certain Calabi-Yau threefold singularities used in geometric
engineering in \cite{KatzVafa}. One starts by considering the
(possibly partial) smoothing of a hyperkahler ADE singularity to a milder
singular
space, parametrized by a triplet of resolution parameters (D-terms or
moment maps in the Hyperk\" ahler construction of the space). The kind of
7-dimensional singularities of interest are obtained by considering a
3-dimensional base parametrizing the resolution parameters, on which one
fibers the corresponding resolved Hyperk\" ahler space. The geometry is
said to be the unfolding of the higher singularity into the lower one.
This construction
guarantees the total geometry admits a $G_2$ holonomy metric. To determine the
matter content arising from the singularity, one decomposes the adjoint
representation of the
A-D-E group associated to the higher singularity with respect to that of the
lower. One obtains chiral fermions with quantum numbers in the
corresponding coset, and multiplicity given by an index which for an
isolates singularity is one.

%The idea is to deform the higher singularity 
%to a lower one (or ``unfolding''), and the
%matter follows from the decomposition of the adjoint representation
%of $G$ where $G$ is the corresponding A-D-E group of the higher
%singularity.

It is easy to realize this construction arises in the M theory lift of the
models presented in the previous section. For example, at points where two stacks of $N$ D6-branes and
$M$ D6-branes intersect, the  M theory  lift corresponds to a singularity
of the $G_2$ holonomy space that represents the unfolding of an $A_{M+N-1}$ singularity into a
4-manifold with an $A_{M_1}$ and an $A_{N-1}$ singularity.  By the
decomposition of the adjoint  representation of $A_{M+N-1}$, we expect the charged
matter to be in the bi-fundamental representation of the $SU(N)\times SU(M)$
gauge group, in agreement with the IIA picture. A different kind of
intersection arises when $N$ D6-branes intersect with an O6-plane, and
consequently with the $N$ D6-brane images. The M theory lift corresponds to
the unfolding of a $D_N$ type  singularity 
into an $A_{N-1}$ singularity.
The decomposition of the adjoint representation  predicts the appearance of chiral
fermions in the antisymmetric representation of $SU(N)$, in agreement with
the IIA picture. In fact this is the origin of the ${\bf 10}$
representations in our previous $SU(5)$ model. 

%\medskip

We  would also  like to note that the generic class of models
 described here may exhibit
some interesting phenomena, {\it e.g.}, the existence
of non-perturbative equivalences among seemingly different models, which
nonetheless share the same M theory lift, in analogy with \cite{kmg}.
On the other hand the type IIA transitions in which intersecting D6-branes
recombine (which are T-dual of small instanton transitions) would have
interesting M theory descriptions, in which the topology of the $G_2$
 holonomy space changes. It would be interesting to explore possible connections of
such process with \cite{AMV,pioline}. We hope that our explicit
constructions may provide a useful laboratory to probe new ideas in
the studies of manifolds with $G_2$ holonomy.

\section*{Acknowledgments}
We thank Gerardo Aldazabal, 
Savas Dimopoulos, Jens Erler,
Gary Gibbons, Jaume Gomis, Luis Ib\'a\~nez, 
Paul Langacker, Hong L\" u,  Chris Pope, Raul Rabad\'an,
Matt Strassler and
Edward Witten for discussions. This work was supported in part by
U.S.\ Department of Energy Grant
No.~DOE-EY-76-02-3071 and
DE-FG02-95ER40893, the Class of 1965 Endowed Term Chair, 
UPenn SAS Dean's funds and the NATO Linkage grant 97061.


\begin{thebibliography}{99}

\bibitem{CSU1}
M.~Cveti\v c, G.~Shiu and A.~M.~Uranga, 
%``Three-family
%supersymmetric standard like models from intersecting  brane
%worlds,'' 
Phys.\ Rev.\ Lett.\  {\bf 87}, 201801 (2001).
%hep-th/0107143.

\bibitem{CSU2}
M.~Cveti\v c, G.~Shiu and A.~M.~Uranga, 
%``Chiral four-dimensional
%N = 1 supersymmetric type IIA orientifolds from  intersecting
%D6-branes,'' 
Nucl.\ Phys.\ B {\bf 615}, 3 (2001).
% [hep-th/0107166]. 

\bibitem{review} For reviews, see, {\it e.g.},
B.~R.~Greene,
{\it Lectures at Trieste Summer School on High Energy Physics and 
Cosmology (1990)};
F.~Quevedo,
%``Lectures on superstring phenomenology,''
{\tt hep-th/9603074};
A.~E.~Faraggi,
%``Superstring phenomenology: Present and future perspective,''
{\tt hep-ph/9707311};
Z.~Kakushadze, G.~Shiu, S.~H.~Tye and Y.~Vtorov-Karevsky,
%``A review of three-family grand unified string models,''
Int.\ J.\ Mod.\ Phys.\ A {\bf 13}, 2551 (1998);
G.~Cleaver, M.~Cveti\v c, J.~R.~Espinosa, L.~L.~Everett, P.~Langacker and
J.~Wang,
Phys.\ Rev.\ D {\bf 59}, 055005 (1999), and references therein.

%\cite{Erler:2001zg}
\bibitem{TypeIInogo}
L.~J.~Dixon, V.~Kaplunovsky and C.~Vafa,
%``On Four-Dimensional Gauge Theories From Type Ii Superstrings,''
Nucl.\ Phys.\ B {\bf 294}, 43 (1987); for generalization of this
theorem to less than 4 dimensions, see,
J.~Erler and G.~Shiu,
%``On type II superstrings in less than four dimensions,''
Phys.\ Lett.\ B {\bf 521}, 114 (2001).
%hep-th/0108230.
 
\bibitem{ABPSS}
C.~Angelantonj, M.~Bianchi, G.~Pradisi, A.~Sagnotti
and Ya.S.~Stanev, \PLB{385}{96}{96}. 

\bibitem{berkooz}
M.~Berkooz and R.G.~Leigh, \NPB{483}{97}{187}.

\bibitem{N1orientifolds}{
Z.~Kakushadze and G.~Shiu, \PRD{56}{97}{3686}; \NPB{520}{98}{75};
Z.~Kakushadze, \NPB{512}{98}{221}.}

\bibitem{zwart}
G.~Zwart, \NPB{526}{98}{378}; D.~O'Driscoll, {\tt hep-th/9801114};

\bibitem{ShiuTye}{G.~Shiu and S.-H.H. Tye, 
Phys. Rev. {\bf D58} (1998) 106007.}

\bibitem{afiv}{G.~Aldazabal, A.~Font, L.E.~Ib\'a\~nez and 
G.~Violero, \NPB{536}{99}{29}.}

\bibitem{wlm} {Z.~Kakushadze, \PLB{434}{98}{269}; 
\PRD{58}{98}{101901};
\NPB{535}{98}{311}.} 

\bibitem{CPW}{M.~Cveti\v c, M.~Pl\" umacher and J.~Wang,
{\it JHEP} {\bf 0004}  (2000) 004.}

\bibitem{kr}
M.~Klein, R.~Rabad\'an, 
%`Z(N) x Z(M) orientifolds with and without discrete torsion' 
JHEP {\bf 0010} (2000) 049.

\bibitem{CUW}{
M.~Cveti\v c, A.~M.~Uranga and J.~Wang,
%`Discrete Wilson lines in N = 1 D = 4 type IIB orientifolds: A
%systematic exploration for Z(6) orientifold,'
Nucl.\ Phys.\ B {\bf 595}, 63 (2001).}

\bibitem{cl}
M.~Cveti\v c and  P.~Langacker,
% `D = 4 N = 1 type IIB orientifolds with
%continuous Wilson lines, moving branes and their field theory realization', 
Nucl.\ Phys.\  {\bf B586} (2000)  287.

\bibitem{sagnottietal}
M.~Bianchi, G.~Pradisi and A.~Sagnotti,
%``Toroidal compactification and symmetry breaking in open string theories,''
Nucl.\ Phys.\ B {\bf 376}, 365 (1992);
M.~Bianchi,
%``A note on toroidal compactifications of the type I superstring and  
%other superstring vacuum configurations with 16 supercharges,''
Nucl.\ Phys.\ B {\bf 528}, 73 (1998)
%, hep-th/9711201;
E.~Witten,
%``Toroidal compactification without vector structure,''
JHEP {\bf 9802}, 006 (1998);
%\bibitem{KST}
Z.~Kakushadze, G.~Shiu and S.-H.~H.~Tye,
%``Type IIB orientifolds with NS-NS antisymmetric tensor backgrounds,''
Phys.\ Rev.\ D {\bf 58}, 086001 (1998).
%hep-th/9803141

\bibitem{bachas}
C.~Bachas, 
%`A Way to break supersymmetry', 
{\tt hep-th/9503030}.

\bibitem{magnetised}
C.~Angelantonj, I.~Antoniadis, E.~Dudas and A.~Sagnotti, 
%`Type I strings on magnetized orbifolds and brane transmutation',
Phys. Lett. B {\bf 489} (2000) 223.

\bibitem{bkl}
R.~Blumenhagen, B.~K\"ors and D.~L\"ust,
%`Type I strings with F flux and B flux',
JHEP {\bf 0102} (2001) 030.

\bibitem{blumen}
R.~Blumenhagen, L.~G\"orlich and B.~K\"ors,
%`Supersymmetric orientifolds in 6-D with D-branes at angles',
Nucl. Phys. B {\bf 569} (2000) 209;
JHEP {\bf 0001} (2000) 040.

\bibitem{bonn}
S.~F\"orste, G.~Honecker and R.~Schreyer,
%`Supersymmetric $\IZ_N \times \IZ_M$ orientifolds in 4-D with D branes at
%angles',
Nucl. Phys. B {\bf 593} (2001) 127; JHEP {\bf 0106} (2001) 004.

\bibitem{bdl}
M.~Berkooz, M.~R.~Douglas and R.~G.~Leigh, 
%`Branes intersecting at angles'
Nucl. Phys. B {\bf 480} (1996) 265.

\bibitem{adsau}
I.~Antoniadis, E.~Dudas and A.~Sagnotti, 
% `Brane supersymmetry breaking', 
Phys. Lett. B {\bf 464} (1999) 38;
G.~Aldazabal and  A.~M.~Uranga, 
% `Tachyon free nonsupersymmetric type IIB orientifolds via
%brane-antibrane systems'
{\it JHEP} {\bf 9910} (1999) 024. 



\bibitem{aiq}
G.~Aldazabal, L.~E.~Ib\'a\~nez and  F.~Quevedo, 
% `Standard - like models with broken supersymmetry from type I string
%vacua'
{\it JHEP} {\bf 0001} (2000) 031, {\tt hep-th/9909172};
% `A D-brane alternative to the MSSM' 
JHEP {\bf 0002} (2000) 015.

\bibitem{aiqu}{
G.~Aldazabal, L.~E.~Ibanez, F.~Quevedo and A.~M.~Uranga,
%``D-branes at singularities: A bottom-up approach to the string
%embedding  of the standard model,''
{\tt hep-th/0005067}}.

\bibitem{bgkl}
R.~Blumenhagen, L.~G\"orlich, B.~K\"ors and D.~L\"ust, 
%`Noncommutative compactifications of type I strings on tori with magnetic
%background flux',
JHEP {\bf 0010} (2000) 006.



\bibitem{afiru}
G.~Aldazabal, S.~Franco, L.~E.~Ib\'a\~nez, R.~Rabad\'an and A.~M.~Uranga,
%`D = 4 chiral string compactifications from intersecting branes' 
Journal of Mathematical Physics, vol. 42, number 7, p. 3103, {\tt
hep-th/0011073}; JHEP {\bf 0102} (2001) 047.


\bibitem{imr}
L.~E.~Ib\'a\~nez, F.~Marchesano and R.~Rabad\'an,
%`Getting just the standard model at intersecting branes' 
{\tt hep-th/0105155}.

\bibitem{bklo}
R.~Blumenhagen, B.~K\"ors and D.~L\"ust, T.~Ott,
%`Noncommutative compactifications of type I strings on tori with magnetic
%background flux',
Nucl. Phys. {\bf B616} (2001) 3.


\bibitem{nsns}
E.~Dudas and J.~Mourad, 
%` Brane solutions in strings with broken supersymmetry and dilaton
% tadpoles', 
Phys. Lett. B {\bf 486} (2000) 172, hep-th/0004165; 
R.~Blumenhagen and A.~Font, 
%`Dilaton tadpoles, warped geometries and large extra dimensions for
%nonsupersymmetric strings', 
Nucl. Phys. B {\bf 599} (2001) 241.

\bibitem{ab}
C.~Angelantonj and R.~Blumenhagen, 
% `Discrete deformations in type I vacua',
Phys. Lett. B {\bf 473} (2000) 86.

\bibitem{small}
E.~Witten,
%``Small Instantons in String Theory,''
Nucl.\ Phys.\ B {\bf 460}, 541 (1996).
\bibitem{cglp}
M.~Cveti\v c, G.~W.~Gibbons, H.~L\" u and C.~N.~Pope,
%``Resolved branes and M-theory on special holonomy spaces,''
hep-th/0106177, String 2001 Proceedings contribution.
%%CITATION = HEP-TH 0106177;%%
\bibitem{acharya}
B.~S.~Acharya,
%``On realising N = 1 super Yang-Mills in M theory,''
hep-th/0011089.

\bibitem{AMV}
M.~Atiyah, J.~Maldacena and C.~Vafa,
%``An M-theory flop as a large N duality,''
hep-th/0011256.


\bibitem{AW}
M.~Atiyah and E.~Witten,
%``M-theory dynamics on a manifold of G(2) holonomy,''
hep-th/0107177.


\bibitem{Witten}
E.~Witten,
%``Anomaly cancellation on G(2)manifolds,''
hep-th/0108165.

\bibitem{aW}
B.~Acharya and E.~Witten,
%``Chiral fermions from manifolds of G(2) holonomy,''
hep-th/0109152.

\bibitem{townsend}
P.~K.~Townsend, Phys. Lett. 350 (1995) 184.

%\cite{Seiberg:1996nz}
\bibitem{SeibergWitten}
N.~Seiberg,
%``IR dynamics on branes and space-time geometry,''
Phys.\ Lett.\ B {\bf 384}, 81 (1996);
N.~Seiberg and E.~Witten,
%``Gauge dynamics and compactification to three dimensions,''
hep-th/9607163.

\bibitem{Sen}
A.~Sen,
%``A note on enhanced gauge symmetries in M- and string theory,''
JHEP {\bf 9709}, 001 (1997).

\bibitem{blumen4d}
R.~Blumenhagen, L.~Gorlich, B.~Kors,
%`Supersymmetric 4-D orientifolds of type IIA with D6-branes at angles',
JHEP {\bf 0001} (2000) 040.

\bibitem{bonn1}
S.~F\"orste, G.~Honecker, R.~Schreyer,
%`Supersymmetric $\IZ_N \times \IZ_M$ orientifolds in 4-D with D branes at
%angles',
Nucl. Phys. B {\bf 593} (2001) 127.

\bibitem{kmg}
S.~Kachru and J.~McGreevy,
%``M-theory on manifolds of G(2) holonomy and type IIA orientifolds,''
JHEP {\bf 0106}, 027 (2001).

\bibitem{KatzVafa}
S.~Katz and C.~Vafa,
%``Matter from geometry,''
Nucl.\ Phys.\ B {\bf 497}, 146 (1997).

\bibitem{pioline}
H.~Partouche and B.~Pioline,
%`Rolling among G(2) vacua',
JHEP {\bf 0103} (2001) 005, {\tt hep-th/0011130}.

\end{thebibliography}
\end{document}